\documentclass[12pt,titlepage,letterpaper]{utarticle}

\usepackage{amsmath}
\usepackage{amsfonts}
\usepackage{amssymb}
\usepackage{amsthm}
\usepackage{xparse}
\usepackage{mathtools}
\usepackage{graphicx}
\usepackage{color}
\usepackage[utf8x]{inputenc}
\usepackage{tocloft}
\usepackage[titletoc,title]{appendix}
\usepackage{cite}
\usepackage{hyperref}
\usepackage{longtable}

\definecolor{aqua}{rgb}{0, 1.0, 1.0}
\definecolor{fuschia}{rgb}{1.0, 0, 1.0}
\definecolor{gray}{rgb}{0.502, 0.502, 0.502}
\definecolor{lime}{rgb}{0, 1.0, 0}
\definecolor{maroon}{rgb}{0.502, 0, 0}
\definecolor{navy}{rgb}{0, 0, 0.502}
\definecolor{olive}{rgb}{0.502, 0.502, 0}
\definecolor{purple}{rgb}{0.502, 0, 0.502}
\definecolor{silver}{rgb}{0.753, 0.753, 0.753}
\definecolor{teal}{rgb}{0, 0.502, 0.502}

\makeatletter
\newdimen\itex@wd%
\newdimen\itex@dp%
\newdimen\itex@thd%
\def\itexspace#1#2#3{\itex@wd=#3em%
\itex@wd=0.1\itex@wd%
\itex@dp=#2ex%
\itex@dp=0.1\itex@dp%
\itex@thd=#1ex%
\itex@thd=0.1\itex@thd%
\advance\itex@thd\the\itex@dp%
\makebox[\the\itex@wd]{\rule[-\the\itex@dp]{0cm}{\the\itex@thd}}}
\makeatother

\makeatletter
\newif\if@sup
\newtoks\@sups
\def\append@sup#1{\edef\act{\noexpand\@sups={\the\@sups #1}}\act}%
\def\reset@sup{\@supfalse\@sups={}}%
\def\mk@scripts#1#2{\if #2/ \if@sup ^{\the\@sups}\fi \else%
  \ifx #1_ \if@sup ^{\the\@sups}\reset@sup \fi {}_{#2}%
  \else \append@sup#2 \@suptrue \fi%
  \expandafter\mk@scripts\fi}
\def\tensor#1#2{\reset@sup#1\mk@scripts#2_/}
\def\multiscripts#1#2#3{\reset@sup{}\mk@scripts#1_/#2%
  \reset@sup\mk@scripts#3_/}
\makeatother

\makeatletter
\newbox\slashbox \setbox\slashbox=\hbox{$/$}
\def\itex@pslash#1{\setbox\@tempboxa=\hbox{$#1$}
  \@tempdima=0.5\wd\slashbox \advance\@tempdima 0.5\wd\@tempboxa
  \copy\slashbox \kern-\@tempdima \box\@tempboxa}
\def\slash{\protect\itex@pslash}
\makeatother

\def\clap#1{\hbox to 0pt{\hss#1\hss}}

\let\oldroot\root
\def\root#1#2{\oldroot #1 \of{#2}}
\renewcommand{\sqrt}[2][]{\oldroot #1 \of{#2}}

\DeclareSymbolFont{symbolsC}{U}{txsyc}{m}{n}
\SetSymbolFont{symbolsC}{bold}{U}{txsyc}{bx}{n}
\DeclareFontSubstitution{U}{txsyc}{m}{n}

\DeclareSymbolFont{stmry}{U}{stmry}{m}{n}
\SetSymbolFont{stmry}{bold}{U}{stmry}{b}{n}

\DeclareFontFamily{OMX}{MnSymbolE}{}
\DeclareSymbolFont{mnomx}{OMX}{MnSymbolE}{m}{n}
\SetSymbolFont{mnomx}{bold}{OMX}{MnSymbolE}{b}{n}
\DeclareFontShape{OMX}{MnSymbolE}{m}{n}{
    <-6>  MnSymbolE5
   <6-7>  MnSymbolE6
   <7-8>  MnSymbolE7
   <8-9>  MnSymbolE8
   <9-10> MnSymbolE9
  <10-12> MnSymbolE10
  <12->   MnSymbolE12}{}

\makeatletter
\def\re@DeclareMathSymbol#1#2#3#4{%
    \let#1=\undefined
    \DeclareMathSymbol{#1}{#2}{#3}{#4}}
\re@DeclareMathSymbol{\neArrow}{\mathrel}{symbolsC}{116}
\re@DeclareMathSymbol{\neArr}{\mathrel}{symbolsC}{116}
\re@DeclareMathSymbol{\seArrow}{\mathrel}{symbolsC}{117}
\re@DeclareMathSymbol{\seArr}{\mathrel}{symbolsC}{117}
\re@DeclareMathSymbol{\nwArrow}{\mathrel}{symbolsC}{118}
\re@DeclareMathSymbol{\nwArr}{\mathrel}{symbolsC}{118}
\re@DeclareMathSymbol{\swArrow}{\mathrel}{symbolsC}{119}
\re@DeclareMathSymbol{\swArr}{\mathrel}{symbolsC}{119}
\re@DeclareMathSymbol{\nequiv}{\mathrel}{symbolsC}{46}
\re@DeclareMathSymbol{\Perp}{\mathrel}{symbolsC}{121}
\re@DeclareMathSymbol{\Vbar}{\mathrel}{symbolsC}{121}
\re@DeclareMathSymbol{\sslash}{\mathrel}{stmry}{12}
\re@DeclareMathSymbol{\bigsqcap}{\mathop}{stmry}{"64}
\re@DeclareMathSymbol{\biginterleave}{\mathop}{stmry}{"6}
\re@DeclareMathSymbol{\invamp}{\mathrel}{symbolsC}{77}
\re@DeclareMathSymbol{\parr}{\mathrel}{symbolsC}{77}
\makeatother

\makeatletter
\def\Decl@Mn@Delim#1#2#3#4{%
  \if\relax\noexpand#1%
    \let#1\undefined
  \fi
  \DeclareMathDelimiter{#1}{#2}{#3}{#4}{#3}{#4}}
\def\Decl@Mn@Open#1#2#3{\Decl@Mn@Delim{#1}{\mathopen}{#2}{#3}}
\def\Decl@Mn@Close#1#2#3{\Decl@Mn@Delim{#1}{\mathclose}{#2}{#3}}
\Decl@Mn@Open{\llangle}{mnomx}{'164}
\Decl@Mn@Close{\rrangle}{mnomx}{'171}
\Decl@Mn@Open{\lmoustache}{mnomx}{'245}
\Decl@Mn@Close{\rmoustache}{mnomx}{'244}
\makeatother

\makeatletter
\DeclareRobustCommand\widecheck[1]{{\mathpalette\@widecheck{#1}}}
\def\@widecheck#1#2{%
    \setbox\z@\hbox{\m@th$#1#2$}%
    \setbox\tw@\hbox{\m@th$#1%
       \widehat{%
          \vrule\@width\z@\@height\ht\z@
          \vrule\@height\z@\@width\wd\z@}$}%
    \dp\tw@-\ht\z@
    \@tempdima\ht\z@ \advance\@tempdima2\ht\tw@ \divide\@tempdima\thr@@
    \setbox\tw@\hbox{%
       \raise\@tempdima\hbox{\scalebox{1}[-1]{\lower\@tempdima\box
\tw@}}}%
    {\ooalign{\box\tw@ \cr \box\z@}}}
\makeatother

\makeatletter
\NewDocumentCommand\mathraisebox{moom}{%
\IfNoValueTF{#2}{\def\@temp##1##2{\raisebox{#1}{$\m@th##1##2$}}}{%
\IfNoValueTF{#3}{\def\@temp##1##2{\raisebox{#1}[#2]{$\m@th##1##2$}}%
}{\def\@temp##1##2{\raisebox{#1}[#2][#3]{$\m@th##1##2$}}}}%
\mathpalette\@temp{#4}}
\makeatletter

\makeatletter
\def\udots{\mathinner{\mkern2mu\raise\p@\hbox{.}
\mkern2mu\raise4\p@\hbox{.}\mkern1mu
\raise7\p@\vbox{\kern7\p@\hbox{.}}\mkern1mu}}
\makeatother

\theoremstyle{plain}

\theoremstyle{definition}

\theoremstyle{remark}

\begin{document}

\preprint{
UTTG--11--17\\
}

\title{Product SCFTs in Class-S}

\author{Jacques Distler, Behzat Ergun and Fei Yan
     \oneaddress{
      Theory Group\\
      Department of Physics,\\
      University of Texas at Austin,\\
      Austin, TX 78712, USA \\
      {~}\\
      \email{distler@golem.ph.utexas.edu}\\
      \email{fei.yan@utexas.edu}\\
      \email{bergun@utexas.edu}\\
      }
}
\date{November 14, 2017}

\Abstract{
We develop a technique for counting the number of stress tensor multiplets in a 4D $\mathcal{N}=2$ SCFT. This provides a simple diagnostic for when an isolated (non-Lagrangian) SCFT is a product of two (or more) such theories. In class-S, the basic building blocks are the isolated SCFTs arising from the compactification of a 6D (2,0) theory on a 3-punctured sphere (``fixture"). We apply our technique to determine when a fixture is a product SCFT. The answer is that this phenomenon is surprisingly rare. In the low-rank $A_{N-1}$, $D_N$ theories and the $E_6$ theory studied by the first author and his collaborators, it occurs less than $1\%$ of the time. Of the 2979 fixtures in the (untwisted and  twisted) $E_6$ theory, only 23 are product SCFTs. Of these, 22 were known to the original authors. The new one is a product of the ${(E_7)}_8$ Minahan-Nemeschansky theory and a new rank-2 SCFT.
}

\maketitle

\tocloftpagestyle{empty}
\tableofcontents
\vfill
\newpage
\setcounter{page}{1}

\section{Introduction}\label{introduction}

In recent years, the class-S construction \cite{Gaiotto:2009hg,Gaiotto:2009we} has yielded a wealth of information about 4D $\mathcal{N}=2$ supersymmetric field theories and their superconformal fixed points. Generically, $\mathcal{N}=2$ SCFTs come in families, where the exactly-marginal deformation corresponds to varying a complex gauge coupling (whose $\beta$-function vanishes). If we turn off the gauge coupling(s), these theories decompose into a product of free vector multiplets with an isolated SCFT, a subgroup of whose global symmetry we had previously gauged.

So, to classify such theories, it suffices to classify the isolated theories and their possible gauging. In class-S, the isolated theories further decompose into products of SCFTs associated to 3-punctured spheres (``fixtures''), on which one performs a partially-twisted compactification of a 6d $(2,0)$ theory. The fixtures fall\footnote{This is not quite true in the twisted compactifications of the $(2,0)$ theories with outer-automorphisms. There \cite{Chacaltana:2012ch,Chacaltana:2013oka,Chacaltana:2015bna,Chacaltana:2016shw}, one encounters a fourth type of fixture, with a hidden marginal deformation, which we called a ``gauge theory fixture.'' Some of these also turn out to be product theories, as we shall see below.}  into three broad types: free hypermultiplets, an isolated \emph{interacting} SCFT, or a mixture of both.

For any \emph{given} $(2,0)$ theory, the list of fixtures is finite, permitting a complete classification of the resulting 4D SCFTs \cite{Tachikawa:2010vg,Chacaltana:2010ks,Chacaltana:2011ze,Chacaltana:2012zy,Chacaltana:2012ch,Chacaltana:2013oka,Chacaltana:2014jba,Chacaltana:2014ica,Chacaltana:2014nya,Chacaltana:2015bna,Chacaltana:2016shw,Chacaltana:2017boe}. It turns out that the \emph{same} isolated 4D SCFT can have many different realizations as fixtures in (different) $(2,0)$ theories. That redundancy is not too difficult to keep track of. More serious is the possibility that some (many?~most?) fixtures could \emph{themselves} correspond to product SCFTs, introducing a further (unexpected) level of redundancy.

This has already been noted, in examples, in \cite{Chacaltana:2011ze,Chacaltana:2012ch,Chacaltana:2014jba,Chacaltana:2015bna,Chacaltana:2017boe}, where the fact that one has a product SCFT can be seen by doing some gauging and then using S-duality (see, e.g., the discussion in \S7 of \cite{Chacaltana:2015bna}). But how prevalent the phenomenon -- of a fixture corresponding to a product SCFT -- is, was unknown.

The purpose of the present paper is to develop a technique for deciding the issue, and applying it to a large (but far from exhaustive) subset of the class-S theories which have been catalogued so-far. The technique will involve using (certain limits of) the superconformal index to compute the number of $\mathcal{N}=2$ stress tensor multiplets (after suitably removing the contribution to the index from any free hypermultiplets that might be present).

For the $A_{N-1}$ and $D_N$ $(2,0)$ theories (at least for low $N$), the number of known product theories is very small. We verify that these are indeed product theories and that there are no additional ones.

We then turn our attention to the $E_6$ $(2,0)$ theory. In the untwisted theory \cite{Chacaltana:2014jba}, there were $10$ fixtures which were known to be products. We checked all 881 fixtures and found no additional product theories. In the twisted sector of the $E_6$ $(2,0)$ theory \cite{Chacaltana:2015bna}, the fixtures were known to include 12 corresponding to product SCFTs. We checked that these were, indeed, product theories and found that there is only one additional previously unknown product theory among the 2078 fixtures in the twisted sector of $E_6$.

From this large, but admittedly still limited sample, we seem to be led to two conclusions.

\begin{itemize}%
\item Fixture that are product SCFTs are relatively rare (at most, a few percent of the total).
\item In all of the examples we have found, whenever you \emph{do} find a product SCFT, one of the factors in the product is always a (rank-$k$) Minahan-Nemeschansky theory \cite{Minahan:1996fg,Minahan:1996cj} (the SCFT whose Higgs branch is the $k$-instanton moduli space of $E_{6,7,8}$). Why this should happen to be the case is a mystery.

\end{itemize}
It would, of course, be of interest to extend this analysis to the much-larger collection of fixtures in the $(2,0)$ theories of type $E_7$ \cite{Chacaltana:2017boe} and $E_8$ \cite{ChacaltanaWIP}. We leave that to future work.

\section{Counting stress tensors}\label{counting_stress_tensors}

The unrefined superconformal index of a 4d $\mathcal{N}=2$ SCFT is defined as \cite{Kinney:2005ej,Gadde:2011uv}:

\begin{displaymath}
I(p,q,t)=\text{Tr}_{\mathcal{H}}(-1)^F p^{\frac{1}{2}(\Delta+2j_1-2R-r)}q^{\frac{1}{2}(\Delta-2j_1-2R-r)}t^{R+r}.
\end{displaymath}
Here $p,q,t$ are the three superconformal fugacities, $\Delta$ is the dilatation generator (conformal Hamiltonian), $j_1$ and $j_2$ are the Cartan generators of the $SU(2)_1\times SU(2)_2$, $R$ and $r$ are the Cartan generators of the $SU(2)_R\times U(1)_r$ R-symmetry. The trace is taken over the Hilbert space $\mathcal{H}$ on $\mathbb{S}^3$ in radial quantization. We will be interested in two specializations of the superconformal index: the Schur index, defined as

\begin{displaymath}
I_{\text{Schur}}= \text{Tr}_{\mathcal{H}}p^{\Delta-R} {(-1)}^F
\end{displaymath}
and the Hall-Littlewood index,

\begin{displaymath}
I_{\text{HL}}= \text{Tr}_{\mathcal{H}_{\text{HL}}}\tau^{2(\Delta-R)} {(-1)}^F
\end{displaymath}
where $\mathcal{H}_{\text{HL}}$ is the subspace of $\mathcal{H}$ defined by $\Delta-2R-r=j_1=0$. The superconformal index does not receive contributions from generic long multiplets of the 4d $\mathcal{N}=2$ superconformal algebra (or from combinations of short multiplets that can recombine into long multiplets).

In the notation of \cite{Dolan:2002zh}, the Hall-Littlewood index receives contributions from the short multiplets $\hat{B}_R$ (whose superconformal primary contributes $\tau^{2R}$) and $D_{R(0,j_2)}$ (whose first superconformal descendent contributes $\tau^{2(R+j_2+1)}{(-1)}^{1+2j_2}$). The Schur index receives contributions from $\hat{C}_{R(j_1,j_2)}$, $\hat{B}_R$, $D_{R,(0,j_2)}$ and $\overline{D}_{R(j_1,0)}$. The contribution from each of these short multiplets is listed in the table below.

\medskip
\begin{tabular}{|l|c|c|}
\hline
Short Multiplet&$I_{\text{Schur}}(p)$&$I_{\text{HL}}(\tau)$\\
\hline 
$\hat{C}_{R(j_1,j_2)}$&$(-1)^{2(j_1+j_2)}\frac{p^{R+j_1+j_2+2}}{1-p}$&$0$\\
\hline
$\hat{B}_R$&$\frac{p^R}{1-p}$&$\tau^{2R}$\\
\hline
$D_{R(0,j_2)}$&$(-1)^{2j_2+1}\frac{p^{R+j_2+1}}{1-p}$&$(-1)^{2j_2+1}\tau^{\frac{R+j_2+1}{2}}$\\
\hline
$\bar{D}_{R(j_1,0)}$&$(-1)^{2j_1+1}\frac{p^{R+j_1+1}}{1-p}$&$0$\\
\hline
\end{tabular}
\medskip

The representation $\hat{B}_{1/2}$ is the free half-hypermultiplet. $D_{0(0,0)}+\overline{D}_{0(0,0)}$ is the free vector multiplet. We assume that there are no free vector multiplets. If there there are free hypermultiplets present, we want to remove their contribution by hand. $n$ free half-hypermultiplets contribute a factor of

\begin{displaymath}
\begin{split}
I_{\text{Schur}} &= {\left(\text{PE}\left[\frac{p^{1/2}}{1-p}\right]\right)}^n = \prod_{k=0}^\infty{\left(\frac{1}{1-p^{k+1/2}}\right)}^n\\
I_{\text{HL}}&= {\left(\text{PE}[\tau]\right)}^n = \frac{1}{{(1-\tau)}^n}
\end{split}
\end{displaymath}
to the index.

After removing the free hypers, we have an isolated interacting SCFT. As such, there should be no higher-spin conserved currents in the spectrum. Various $D_{R,(0,j_2)}$ and $\overline{D}_{R(j_1,0)}$ multiplets contain such higher spin currents and hence must be absent from the spectrum. In particular,

\begin{displaymath}
\begin{gathered}
\#D_{1/2(0,1/2)}=\#\overline{D}_{1/2(1/2,0)}=\#D_{0(0,1)}=\#\overline{D}_{0(1,0)}
=\#D_{1/2(0,0)}\\=\#\overline{D}_{1/2(0,0)}=\#D_{0(0,1/2)}=\#\overline{D}_{0(1/2,0)}=0
\end{gathered}
\end{displaymath}
The remaining contributions to the Schur and Hall-Littlewood indices can be written as follows

\begin{displaymath}
\begin{split}
I_{\text{Schur}} &= 1+s_1 p +s_{3/2}p^{3/2} + s_2 p^2 +\dots\\
I_{\text{HL}}&=1+ h_1\tau^2 + h_{3/2}\tau^3 + h_2 \tau^4+\dots 
\end{split}
\end{displaymath}
where

\begin{displaymath}
\begin{split}
h_1&=s_1=\#\hat{B}_1\\
h_{3/2}&=s_{3/2}=\#\hat{B}_{3/2}\\
h_2&= \#\hat{B}_2 -\# D_{1(0,0)}\\
s_2&=\#\hat{B}_1+\#\hat{B}_2-\# D_{1(0,0)} - \# \overline{D}_{1(0,0)}+\#\hat{C}_{0(0,0)}
\end{split}
\end{displaymath}
Rearranging these, we obtain

\begin{displaymath}
\#\hat{C}_{0(0,0)}=s_2-h_1-h_2+\# \overline{D}_{1(0,0)}
\end{displaymath}
In general, this gives us only a lower bound

\begin{equation}
\#\hat{C}_{0(0,0)}\geq s_2-h_1-h_2
\label{lowerbound}\end{equation}
Because of the recombination formula,

\begin{displaymath}
\hat{C}_{0(0,0)}+D_{1(0,0)}+\overline{D}_{1(0,0)}+\hat{B}_2 = \text{long multiplet}
\end{displaymath}
the superconformal index cannot do better than this lower bound. We need some dynamical information. The key point is that $D_{1(0,0)}+\overline{D}_{1(0,0)}$ is the multiplet containing an $(\mathcal{N}=1)$-preserving (but $(\mathcal{N}=2)$-breaking) marginal perturbation (exactly-marginal, if it's a flavour singlet \cite{Green:2010da}). If such an operator \emph{is} present in our product theory, then one of the factors in the product is actually a special point of enhanced $\mathcal{N}=2$ superconformal symmetry in a \emph{family} of $\mathcal{N}=1$ superconformal theories. While this is certainly possible, it seems unlikely in the cases at hand. So we will simply assume that $\# \overline{D}_{1(0,0)}=0$ and \eqref{lowerbound} is an \emph{equality}. $\hat{C}_{0(0,0)}$ is the $\mathcal{N}=2$ stress tensor multiplet and computing the RHS of \eqref{lowerbound} allows us to count them.

\section{Superconformal Index for Class-S Theories}\label{superconformal_index_for_class-s}

In this section, we'll recall some facts about class-S theories and their superconformal indices. A class-S theory of type $\mathfrak{j}$ is obtained by a partially-twisted compactification of a 6d $(2,0)$ theory of type $\mathfrak{j}$, where $\mathfrak{j}$ is a simply-laced Lie algebra, on a genus-$g$, $n$-punctured Riemann surface $\mathcal{C}_{g,n}$. The punctures are the locations of codimension-2 defects and are labelled by nilpotent orbits in $\mathfrak{j}$ or, equivalently, embeddings $\rho: \mathfrak{su}(2)\rightarrow \mathfrak{j}$ up to conjugation. The global symmetry associated to a puncture is then the centralizer $\mathfrak{f}$ of $\rho(\mathfrak{su}(2))\subseteq \mathfrak{j}$ \cite{Chacaltana:2012zy}.

For a fixture, i.e. a 3-punctured sphere, the Schur and Hall-Littlewood limits of the unrefined superconformal indices have the following form \cite{Gadde:2011uv,Lemos:2012ph}

\begin{align}
	I_{Schur}(p)&= \sum_\Lambda \left.\frac{\prod_{i=1}^3 \mathcal{K}_S(\textbf{a}_i)\chi_\Lambda(\textbf{a}_i)}{\mathcal{K}_S(\{p\})\chi_\Lambda(\{p\})} \right|_{\textbf{a}_i\rightarrow 1}\\
	I_{HL}(\tau)&= \sum_\Lambda \left. \frac{\prod_{i=1}^3 \mathcal{K}_{HL}(\textbf{a}_i)P_\Lambda(\textbf{a}_i)}{\mathcal{K}_{HL}(\{\tau\})P_\Lambda(\{\tau\})}\right|_{\textbf{a}_i\rightarrow 1}
\end{align}
 
where

\begin{enumerate}
	
	\item The sum is over highest weights $\Lambda$ labeling the finite dimensional irreducible representations of $\mathfrak{j}$.
	
	\item Flavor fugacities $\textbf{a}_i$ associated to the $i^{\text{th}}$ puncture are determined by decomposition of the fundamental representation of $\mathfrak{j}$ as a representation of $\rho_i(\mathfrak{su}(2))\times \mathfrak{f}_i$. There's some freedom in assigning these but the choices are equivalent under the action of the Weyl group $W$ of $\mathfrak{j}$. $\{p\}$ and $\{\tau\}$ are the fugacities for the trivial puncture.
	
	\item The $\mathcal{K}$-factor associated to the $i^{\text{th}}$ puncture is determined by the restriction of the adjoint representation $\text{ad}_\mathfrak{j}$ of $\mathfrak{j}$ to $\rho_i(\mathfrak{su}(2))\times \mathfrak{f}_i$ as
	
	\begin{align}
	\text{ad}_{\mathfrak{j}}= \bigoplus_n V_n \otimes R_{n,i}
	\end{align}
	
	where $V_n$ is the $n$-dimensional irreducible representation of $\mathfrak{su}(2)$ and $R_{n,i}$ is the corresponding representation of $\mathfrak{f}_i$, possibly reducible. Upon this decomposition, the $\mathcal{K}$-factors are
	
	\begin{align}
		\mathcal{K}_S(\textbf{a}_i)&= \text{PE}\left[\sum_n \frac{p^{\frac{n+1}{2}}}{1-p} \chi^{\mathfrak{f}_i}_{R_{n,i}}(\textbf{a}_i)\right]\\
		\mathcal{K}_{HL}(\textbf{a}_i)&= (1-\tau^2)^{\frac{\text{rank}(\mathfrak{j})}{2}}\text{PE}\left[\sum_n \tau^{n+1} \chi^{\mathfrak{f}_i}_{R_{n,i}}(\textbf{a}_i)\right]
		\end{align}
	
	\item The polynomials appearing in the index $\chi_\Lambda$ and $P_\Lambda$ are characters and Hall-Littlewood polynomials for the representation labeled by $\Lambda$ respectively. The formula for HL polynomials is
	
	\begin{align}
		P_\Lambda (\textbf{a}_i) &=\frac{1}{W_{\Lambda}(\tau)} \sum_{w\in W} e^{w(\Lambda )} \prod_{\alpha \in \Phi_+}\frac{1- \tau^2 e^{- w(\alpha)}}{1-e^{-w(\alpha)}}\\
		W_\Lambda(\tau)&= \sqrt{\sum_{w \in \text{Stab}_W(\Lambda)}\tau^{2l(w)}}
\end{align}
where $\Phi_+$ are the positive roots of $\mathfrak{j}$ and flavor fugacities $\{\textbf{a}_i\}$ can be assigned once we choose a basis for the weight lattice for $\mathfrak{j}$. 
\end{enumerate}

In the twisted sector, some of the defects might have the action of an outer automorphism $o\in Out(\mathfrak{j})$. Let $\mathfrak{g}\subset\mathfrak{j}$ be the invariant subalgebra. Twisted defects are labeled by, up to conjugation, homomorphisms $\rho: \mathfrak{su}(2)\rightarrow \mathfrak{g}^{\vee}$ where $\mathfrak{g}^{\vee}$ is Langlands dual of $\mathfrak{g}$. As in the untwisted case, the flavor symmetry is the centralizer of the image of $\rho$  \cite{Chacaltana:2012zy}. Twisted-sector fixtures have 2 twisted punctures and 1 untwisted puncture. Unrefined superconformal indices for such fixtures have almost the same form as before but are slightly modified as \cite{Lemos:2012ph,Chacaltana:2013oka,Chacaltana:2015bna}

\begin{align}
	I_{Schur}(p)&= \sum_{\Lambda'} \left.\frac{\mathcal{K}_S(\textbf{b})\chi^{\mathfrak{j}}_{\Lambda}(\textbf{b})
	\prod_{i=2}^3 \bar{\mathcal{K}}_S(\textbf{a}_i)\chi^{\mathfrak{g}^\vee}_{\Lambda'}(\textbf{a}_i) }{\mathcal{K}_S(\{p\})\chi^{\mathfrak{j}}_{\Lambda}(\{p\})} \right|_{\textbf{a}_i,\textbf{b}\rightarrow 1}\\
	I_{HL}(\tau)&= \sum_{\Lambda'} \left. \frac{\mathcal{K}_{HL}(\textbf{b})P^{\mathfrak{j}}_\Lambda(\textbf{b})\prod_{i=2}^3 \bar{\mathcal{K}}_{HL}(\textbf{a}_i)P^{\mathfrak{g}^\vee}_{\Lambda'}(\textbf{a}_i)}{\mathcal{K}_{HL}(\{\tau\})P^{\mathfrak{j}}_{\Lambda}(\{\tau\})}\right|_{\textbf{a}_i,\textbf{b}\rightarrow 1}
\end{align}
where the sum is now over the weights $\Lambda'$ of $\mathfrak{g}^{\vee}$, extended\footnote{For the main case of interest here, namely $\mathfrak{g}^\vee=\mathfrak{f}_4$ and $\mathfrak{j}=\mathfrak{e}_6$, the precise extension can be found in \S4.1 of \cite{Chacaltana:2015bna}.} (in the case of the untwisted puncture) to weights of $\mathfrak{j}$ (denoted as $\Lambda$ in the formulas). The $\bar{\mathcal{K}}$ and flavor fugacities $\textbf{a}_i$ for twisted punctures are determined as in the untwisted case but with $\mathfrak{j}$ replaced by $\mathfrak{g}^\vee$.

The main computational bottleneck is computing and evaluating the Hall-Littlewood polynomials, which requires a sum over the elements of the Weyl group. For low rank classical algebras $A_N$ and $D_N$, the Weyl groups are rather small and the HL polynomials can be evaluated with ease. However, $|W_{E_6}|=51840$, which makes the evaluation of HL polynomials very tedious. And we need to compute them for every representation that contributes to a given order in $\tau$. Fortunately, one can exploit the freedom in the choice of flavor fugacity assignments to deduce whether or not a given representation will contribute to a desired order. 

For the untwisted $E_6$ theory, it turns out there are 71 representations that contribute to the order $p^2$ and $\tau^4$. The highest dimensional representation that occur has Dynkin labels $[0,0,1,0,0,2]_{\mathfrak{e}_6}$ and dimension $=1911195$. In the twisted $E_6$ case, there are 30 representations that contribute, $15$ of which already appeared in the untwisted case. The largest $\mathfrak{f}_4$ and $\mathfrak{e}_6$ representations that appeared to order $p^2$ and $\tau^4$ have $\text{dim}[1,1,0,1]_{\mathfrak{f}_4}=379848$ and $\text{dim}[2,1,0,1,2,0]_{\mathfrak{e}_6}=688740975$.

\section{Examples}\label{examples}

As a simple example, consider the interacting fixture

\begin{displaymath}
 \includegraphics[width=108pt]{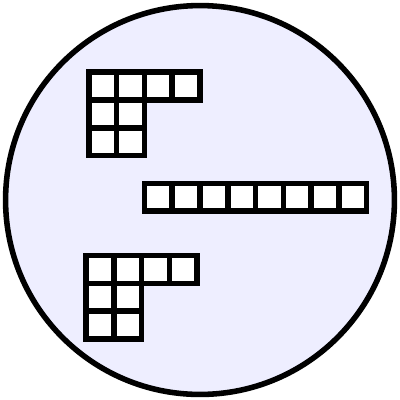}
\end{displaymath}
in the $D_4$ theory. The corresponding 4d $\mathcal{N}=2$ SCFT was identified as the product of two copies of rank-$1$ Minahan-Nemeschansky $E_6$ SCFT in \cite{Chacaltana:2011ze}. The unrefined Schur and Hall-Littlewood indices for this fixture to the order of $p^2$ ($\tau^4$) are

\begin{displaymath}
\begin{split}
I_{\text{Schur}} &= 1+156 p + 11102 p^2+\dots \\
I_{\text{HL}}&=1+ 156\tau^2 + 10944 \tau^4+\dots
\end{split}
\end{displaymath}
We read off $h_1=s_1=156$, which equals the dimension of $\mathfrak{e}_6\oplus\mathfrak{e}_6$. The lower bound on $\#\hat{C}_{0(0,0)}$,

\begin{displaymath}
s_2-h_1-h_2=11102-156-10944=2
\end{displaymath}
is clearly saturated in this example.

At least for low $N$, there are not too many further examples of product SCFTs among the (twisted or untwisted) fixtures of the $A_N$ or $D_N$ theories. For most of the interacting fixtures the lower bound on $\#\hat{C}_{0(0,0)}$ is equal to $1$. However, there are more interesting product SCFTs in theories of type $E_6$.

In the untwisted $E_6$ case, our results can be summarized in the table below. We find 10 product theories among the 881 good fixtures (the numbering is the one used in \cite{Chacaltana:2014jba}) with regular punctures. The first 7 were known to be product theories in \cite{Chacaltana:2014jba}. The last 3 were not.

Those three fixtures,
\begin{displaymath}
\begin{matrix} \includegraphics[width=63pt]{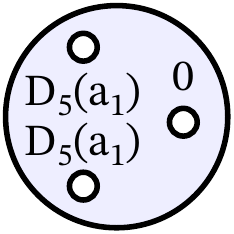}\end{matrix},\qquad
\begin{matrix} \includegraphics[width=63pt]{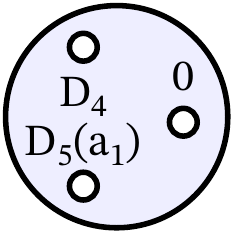}\end{matrix},\qquad
\begin{matrix} \includegraphics[width=63pt]{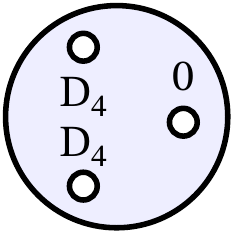}\end{matrix}
\end{displaymath}
(respectively, $\#59$, $61$ and $99$ in the table of interacting fixtures in \cite{Chacaltana:2014jba}) were later identified as product theories in \cite{Chacaltana:2015bna} by gauging a subgroup of the flavour symmetry and using S-duality.

{\footnotesize

\begin{longtable}{|c|c|l|l|c|c|}
\hline
$\#_{type}$&Fixture&$I_{\text{Schur}}(p)$&$I_{\text{HL}}(\tau)$&$\#\hat{C}_{0(0,0)}$&Theory\\
\hline
\endhead
$1_{\text{int}}$&$\begin{matrix} \includegraphics[width=63pt]{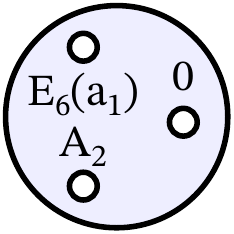}\end{matrix}$&$\begin{aligned}1&+496p+\\&+116002p^2+\ldots\end{aligned}$&$\begin{aligned}1&+496\tau^2+\\&+115504\tau^4+\ldots\end{aligned}$&2&${[{(E_8)}_{12}\,\text{SCFT}]}^2$\\
\hline
$8_{\text{int}}$&$\begin{matrix} \includegraphics[width=63pt]{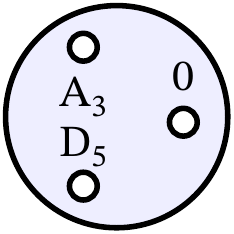}\end{matrix}$&$\begin{aligned}1&+222p+216p^{\frac{3}{2}}+\\&+23880p^2+\ldots\end{aligned}$&$\begin{aligned}1&+222\tau^2+216\tau^3+\\&+23656\tau^4+\ldots\end{aligned}$&2&${\begin{gathered} [{(E_7)}_8\,\text{SCFT}]\\ \times \\ [{(E_6)}_{16} \times {Sp(2)}_{10} \times U(1)\,\text{SCFT}]\end{gathered}}$\\
\hline
$6_{\text{int}}$&$\begin{matrix} \includegraphics[width=63pt]{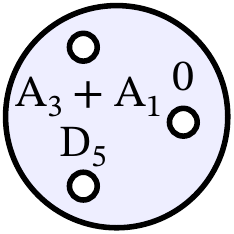}\end{matrix}$&$\begin{aligned}1&+269p+266p^{\frac{3}{2}}+\\&+35045p^2+\ldots\end{aligned}$&$\begin{aligned}1&+269\tau^2+266\tau^3+\\&+34774\tau^4+\ldots\end{aligned}$&2&${\begin{gathered} [{(E_7)}_8\,\text{SCFT}]\\ \times \\ [{(E_7)}_{16} \times {SU(2)}_{9}\,\text{SCFT}]\end{gathered}}$\\
\hline
$39_{\text{int}}$&$\begin{matrix} \includegraphics[width=63pt]{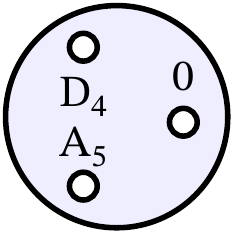}\end{matrix}$&$\begin{aligned}1&+329p+156p^{\frac{3}{2}}+\\&+50739p^2+\ldots\end{aligned}$&$\begin{aligned}1&+329\tau^2+156\tau^3+\\&+50408\tau^4+\ldots\end{aligned}$&2&${\begin{gathered} [{(E_8)}_{12}\,\text{SCFT}]\\ \times \\ [{(E_6)}_{12} \times {SU(2)}_{7}\,\text{SCFT}]\end{gathered}}$\\
\hline
$11_{\text{mix}}$&$\begin{matrix} \includegraphics[width=63pt]{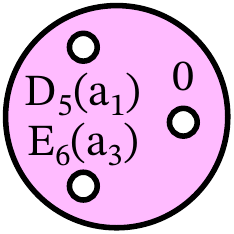}\end{matrix}$&$\begin{aligned}1&+54p^{\frac{1}{2}}+1641p+\\&+36198p^{\frac{3}{2}}+\\&+640688p^2+\ldots\end{aligned}$&$\begin{aligned}1&+54\tau+1641\tau^2+\\&+36144\tau^3+\\&+637614\tau^4+\ldots\end{aligned}$&2&${[{(E_6)}_{6}\,\text{SCFT}]}^2 + 1(27)$\\
\hline
$5_{\text{int}}$&$\begin{matrix} \includegraphics[width=63pt]{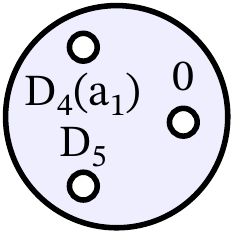}\end{matrix}$&$\begin{aligned}1&+399p+\\&+75582p^2+\ldots\end{aligned}$&$\begin{aligned}1&+399\tau^2+\\&+75180\tau^4+\ldots\end{aligned}$&3&${[{(E_7)}_{8}\,\text{SCFT}]}^3$\\
\hline
$18_{\text{int}}$&$\begin{matrix} \includegraphics[width=63pt]{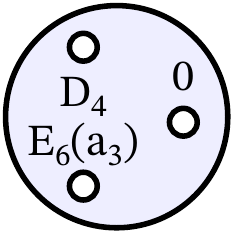}\end{matrix}$&$\begin{aligned}1&+404p+\\&+77039p^2+\ldots\end{aligned}$&$\begin{aligned}1&+404\tau^2+\\&+76632\tau^4+\ldots\end{aligned}$&3&${\begin{gathered} [{(E_8)}_{12}\,\text{SCFT}]\\ \times \\ {[{(E_6)}_{6}\,\text{SCFT}]}^2\end{gathered}}$\\
\hline
$99_{\text{int}}$&$\begin{matrix} \includegraphics[width=63pt]{D4D40}\end{matrix}$&$\begin{aligned}1&+172p+\\&+14886p^2+\ldots\end{aligned}$&$\begin{aligned}1&+172\tau^2+\\&+14712\tau^4+\ldots\end{aligned}$&2&${\begin{gathered} [{(E_6)}_6\,\text{SCFT}]\\ \times \\ [{(E_6)}_{18} \times {SU(3)}_{12}^2\,\text{SCFT}]\end{gathered}}$\\
\hline
$61_{\text{int}}$&$\begin{matrix} \includegraphics[width=63pt]{D4D5a10}\end{matrix}$&$\begin{aligned}1&+165p+164p^{\frac{3}{2}}+\\&+13451p^2+\ldots\end{aligned}$&$\begin{aligned}1&+165\tau^2+164\tau^3+\\&+13284\tau^4+\ldots\end{aligned}$&2&${\begin{gathered} [{(E_6)}_6\,\text{SCFT}]\\ \times \\ [{(E_6)}_{18} \times {SU(3)}_{12} \times U(1)\,\text{SCFT}]\end{gathered}}$\\
\hline
$59_{\text{int}}$&$\begin{matrix} \includegraphics[width=63pt]{D5a1D5a10}\end{matrix}$&$\begin{aligned}1&+212p+112p^{\frac{3}{2}}+\\&+22273p^2+\ldots\end{aligned}$&$\begin{aligned}1&+212\tau^2+112\tau^3+\\&+22059\tau^4+\ldots\end{aligned}$&2&${\begin{gathered} [{(E_6)}_6\,\text{SCFT}]\\ \times \\ [{(E_7)}_{18}\times U(1)\,\text{SCFT}]\end{gathered}}$\\
\hline
\end{longtable}

}

In the twisted $E_6$ case, we identify 13 product theories among 2078 good fixtures with regular punctures. Only one interacting fixture, namely fixture $\#91$, was not previously listed in \cite{Chacaltana:2015bna} as a product theory. We also find that three gauge theory fixtures are product theories. One was explicitly noted as such in \S3.6 of \cite{Chacaltana:2015bna}. We discuss the other two below. Our results can be summarized in the following table.

{\footnotesize

\begin{longtable}{|c|c|l|l|c|c|}
\hline
$\#_{type}$&Fixture&$I_{\text{Schur}}(p)$&$I_{\text{HL}}(\tau)$&$\#\hat{C}_{0(0,0)}$&Theory\\
\hline
\endhead
$111_{\text{int}}$&$\begin{matrix} \includegraphics[width=64pt]{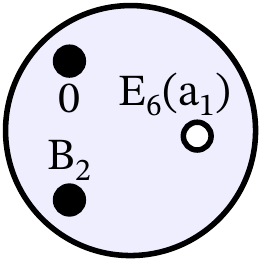}\end{matrix}$&$\begin{aligned}1&+136p+104p^\frac{3}{2}+\\&+9036p^2+\ldots\end{aligned}$&$\begin{aligned}1&+136 \tau^2+104\tau^3+\\&+8898\tau^4+\ldots\end{aligned}$&2&${\begin{gathered} [{(E_6)}_6\,\text{SCFT}]\\ \times \\ [{(F_4)}_{12} \times {SU(2)}_7^2\,\text{SCFT}]\end{gathered}}$\\
\hline
$103_{\text{int}}$&$\begin{matrix} \includegraphics[width=64pt]{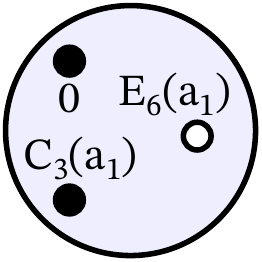}\end{matrix}$&$\begin{aligned}1& + 159 p + 156 p^\frac{3}{2}+\\& + 12229 p^2+\ldots\end{aligned}$&$\begin{aligned}1&+159 \tau^2+156\tau^3+\\&+12068\tau^4+\ldots\end{aligned}$&2&${\begin{gathered} [{(E_6)}_6\,\text{SCFT}]\\ \times \\ [{(E_6)}_{12}\times {SU(2)}_7\,\text{SCFT}]\end{gathered}}$\\
\hline
$99_{\text{int}}$&$\begin{matrix} \includegraphics[width=64pt]{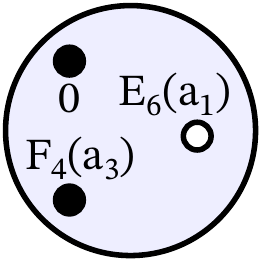}\end{matrix}$&$\begin{aligned}1& + 234 p+\\&  + 25779 p^2+\ldots\end{aligned}$&$\begin{aligned}1&+234 \tau^2+\\&+25542\tau^4+\ldots\end{aligned}$&3&${[{(E_6)}_6 \, \text{SCFT}]}^3$\\
\hline
$91_{\text{int}}$&$\begin{matrix} \includegraphics[width=64pt]{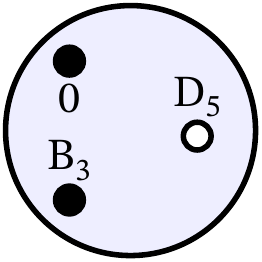}\end{matrix}$&$\begin{aligned}1& + 186 p+\\&+ 16142 p^2+\ldots\end{aligned}$&$\begin{aligned}1&+186 \tau^2+\\&+15954\tau^4+\ldots\end{aligned}$&2&${\begin{gathered} [{(E_7)}_8\,\text{SCFT}]\\ \times \\ [{(F_4)}_{10}\times U(1)\,\text{SCFT}]\end{gathered}}$\\
\hline
$14_{\text{int}}$&$\begin{matrix} \includegraphics[width=64pt]{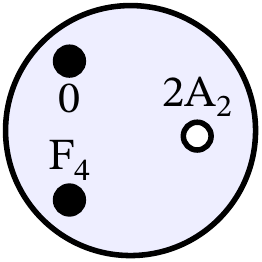}\end{matrix}$&$\begin{aligned}1& + 326 p+\\&+ 49102 p^2+\ldots\end{aligned}$&$\begin{aligned}1&+326 \tau^2+\\&+48774\tau^4+\ldots\end{aligned}$&2&${\begin{gathered} [{(E_8)}_{12}\,\text{SCFT}]\\ \times \\ [{(E_6)}_{6}\,\text{SCFT}]\end{gathered}}$\\
\hline
$5_{\text{int}}$&$\begin{matrix} \includegraphics[width=64pt]{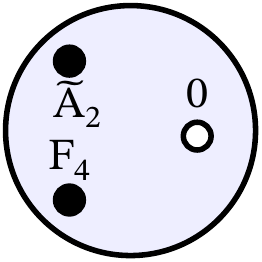}\end{matrix}$&$\begin{aligned}1& + 170 p+\\&+ 14601 p^2+\ldots\end{aligned}$&$\begin{aligned}1&+170 \tau^2+\\&+14429\tau^4+\ldots\end{aligned}$&2&${\begin{gathered} [{(E_6)}_6\,\text{SCFT}]\\ \times \\ [{(E_6)}_{18}\times {(G_2)}_{10}\,\text{SCFT}]\end{gathered}}$\\
\hline
$4_{\text{int}}$&$\begin{matrix} \includegraphics[width=64pt]{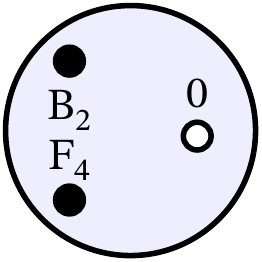}\end{matrix}$&$\begin{aligned}1& + 162 p + 312 p^\frac{3}{2}+\\& + 13365 p^2+\ldots\end{aligned}$&$\begin{aligned}1&+162 \tau^2+312\tau^3+\\&+13201\tau^4+\ldots\end{aligned}$&2&${[{(E_6)}_{12}\times {SU(2)}_7 \,\text{SCFT}]}^2$\\
\hline
$3_{\text{int}}$&$\begin{matrix} \includegraphics[width=64pt]{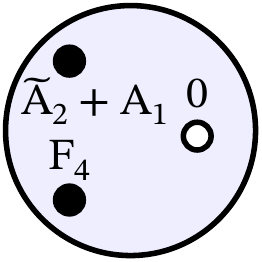}\end{matrix}$&$\begin{aligned}1& + 159 p + 160 p^\frac{3}{2}+\\& + 12464 p^2+\ldots\end{aligned}$&$\begin{aligned}1&+159 \tau^2+160\tau^3+\\&+12303\tau^4+\ldots\end{aligned}$&2&${\begin{gathered} [{(E_6)}_6\,\text{SCFT}]\\ \times \\ [{(E_6)}_{18}\times {SU(2)}_{20}\,\text{SCFT}]\end{gathered}}$\\
\hline
$2_{\text{int}}$&$\begin{matrix} \includegraphics[width=64pt]{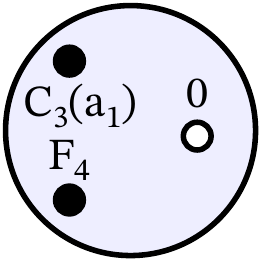}\end{matrix}$&$\begin{aligned}1& + 237 p + 156 p^\frac{3}{2}+\\& + 27140 p^2+\ldots\end{aligned}$&$\begin{aligned}1&+237 \tau^2+156\tau^3+\\&+26900\tau^4+\ldots\end{aligned}$&3&${\begin{gathered} {[{(E_6)}_6\,\text{SCFT}]}^2\\ \times \\ [{(E_6)}_{12}\times {SU(2)}_7\,\text{SCFT}]\end{gathered}}$\\
\hline
$1_{\text{int}}$&$\begin{matrix} \includegraphics[width=64pt]{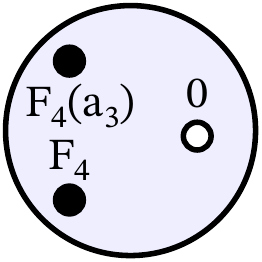}\end{matrix}$&$\begin{aligned}1& + 312 p+\\&+ 46540 p^2+\ldots\end{aligned}$&$\begin{aligned}1&+312 \tau^2+\\&+46224\tau^4+\ldots\end{aligned}$&4&${[{(E_6)}_6 \,\text{SCFT}]}^4$\\
\hline
$n/a_{\text{gauge}}$&$\begin{matrix} \includegraphics[width=64pt]{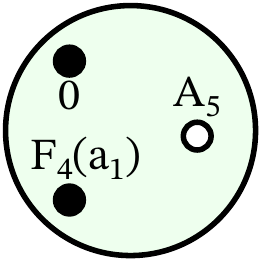}\end{matrix}$&$\begin{aligned}1& + 133 p + 52 p^\frac{3}{2}+\\& + 8446 p^2+\ldots\end{aligned}$&$\begin{aligned}1&+133 \tau^2+52\tau^3+\\&+8311\tau^4+\ldots\end{aligned}$&2&\\
\hline
$n/a_{\text{gauge}}$&$\begin{matrix} \includegraphics[width=64pt]{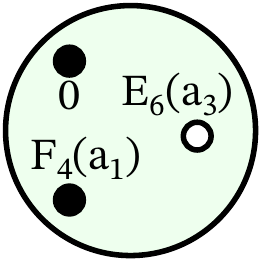}\end{matrix}$&$\begin{aligned}1& + 156 p+\\&+ 11830 p^2+\ldots\end{aligned}$&$\begin{aligned}1&+156 \tau^2+\\&+11672\tau^4+\ldots\end{aligned}$&2&\\
\hline
$2_{\text{gauge}}$&$\begin{matrix} \includegraphics[width=64pt]{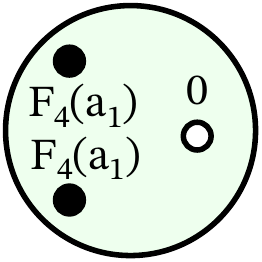}\end{matrix}$&$\begin{aligned}1& + 326 p+\\&+ 12558 p^2+\ldots\end{aligned}$&$\begin{aligned}1&+326 \tau^2+\\&+12400\tau^4+\ldots\end{aligned}$&2&\\
\hline
\end{longtable}
}

\section{Gauge theory fixtures}\label{gauge_theory_fixtures}

The $\underline{F_4(a_1)}$ puncture, in the twisted sector of the $E_6$ theory, is ``atypical'' (in the nomenclature of \cite{Chacaltana:2012ch}). That is, it carries a ``hidden'' marginal deformation. To access the full space of marginal couplings, we should resolve it to a pair of punctures: $\underline{F_4}$ (the simple puncture from the twisted sector) and $E_6(a_1)$ (the simple puncture from the untwisted sector). The coincident limit of those two punctures does not imply any gauge coupling becoming weak; instead, we simply obtain $\underline{F_4(a_1)}$.

A fixture with an $\underline{F_4(a_1)}$ puncture is thus, really, a 4-punctured sphere in disguise:

\begin{displaymath}
 \includegraphics[width=213pt]{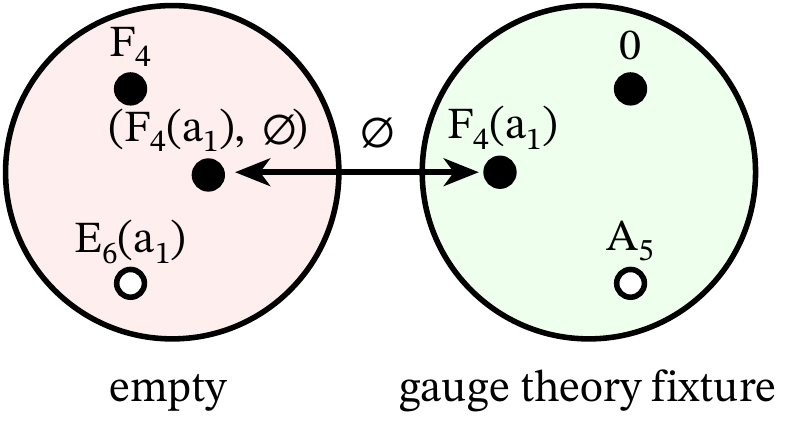}
\end{displaymath}
where the gauge theory is at a strong coupling point in the interior of the conformal manifold. We computed that the theory has two stress tensors, and is thus a product SCFT. That is indeed the case, as we can see by examining the other degenerations of the 4-punctured sphere which is its resolution:

\begin{displaymath}
 \includegraphics[width=275pt]{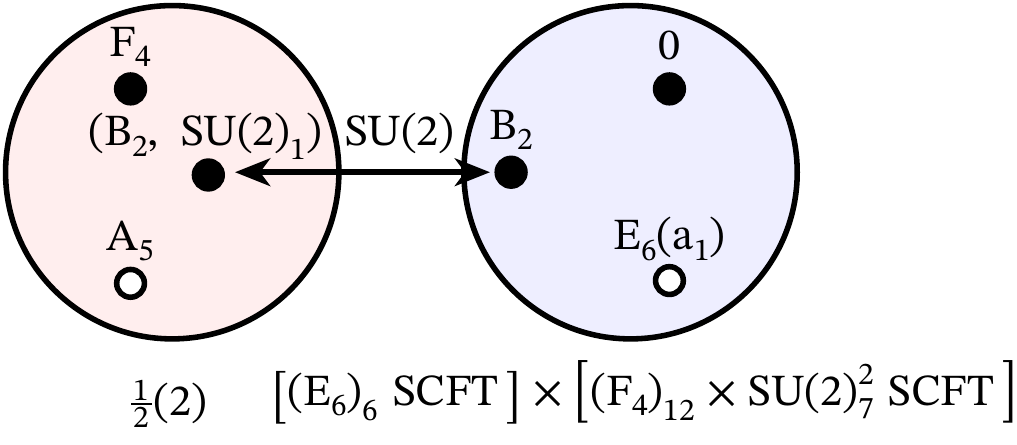}
\end{displaymath}
where one of the $SU(2)$s of the ${(F_4)}_{12}\times {SU(2)}_7^2$ SCFT is gauged, and

\begin{displaymath}
 \includegraphics[width=219pt]{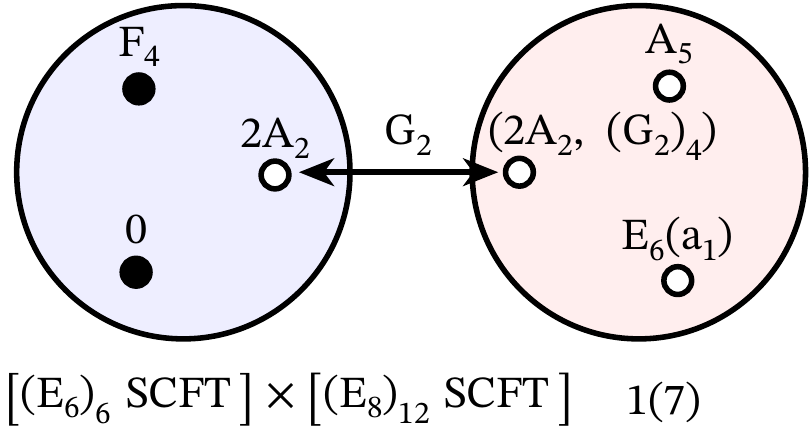}
\end{displaymath}
where a $G_2$ subgroup of the $E_8$ is gauged. In each case, there is a decoupled Minahan-Nemeschansky ${(E_6)}_6$ SCFT, as anticipated.

The same remarks apply, \emph{mutatis mutandis}, to

\begin{displaymath}
 \includegraphics[width=213pt]{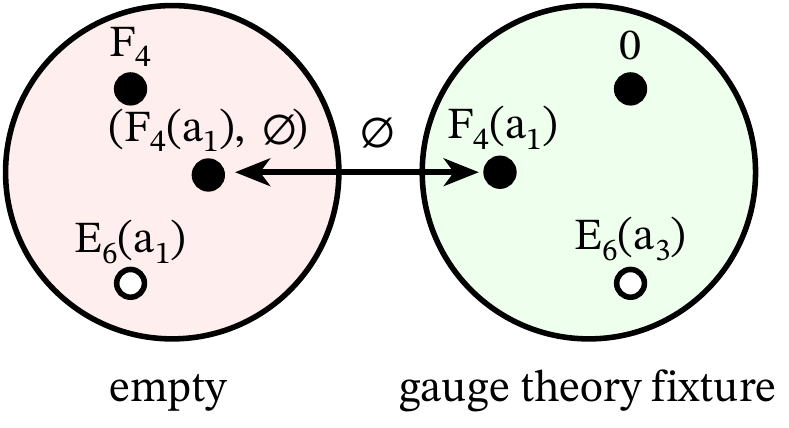}
\end{displaymath}
(see \S8.1 of \cite{Chacaltana:2015bna}, where this and the third gauge-theory fixture are discussed in detail) whose S-dual frames are

\begin{displaymath}
 \includegraphics[width=275pt]{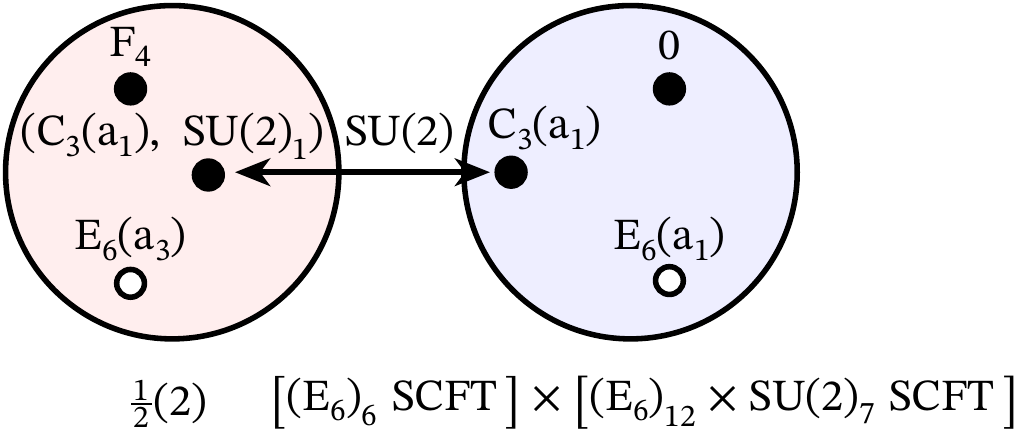}
\end{displaymath}
and

\begin{displaymath}
 \includegraphics[width=219pt]{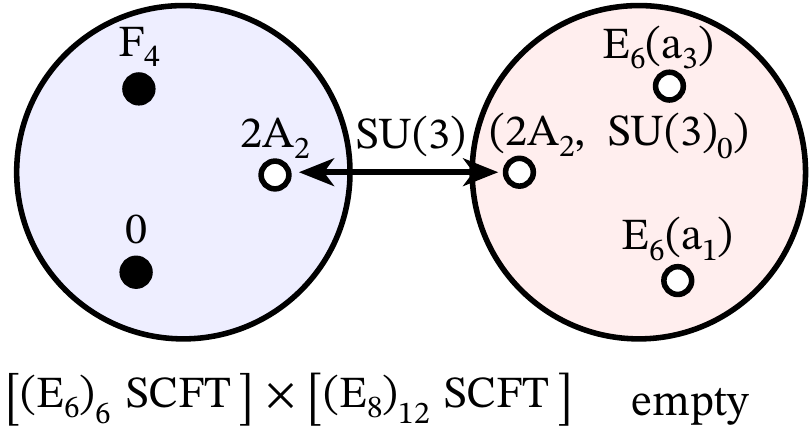}
\end{displaymath}
Here, too, there is a decoupled ${(E_6)}_6$ SCFT.

This is a nice check that our formalism works, even when the SCFTs are not isolated.

\section{The new product SCFT}\label{the_new_product_scft}

The fixture
\begin{displaymath}
 \includegraphics[width=189pt]{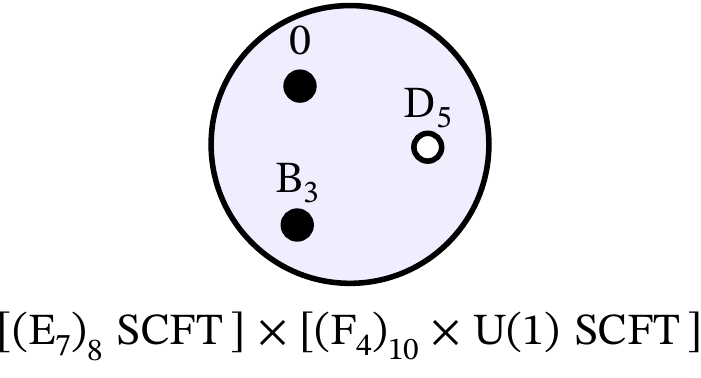}
\end{displaymath}
is a product SCFT that hasn't been identified previously\footnote{Only an ${(F_4)}_{18}\times {SU(2)}_{24}\times U(1)$ subgroup of the global symmetry is manifest. Of this, only (a subgroup of the)  ${(F_4)}_{18}\subset {(E_7)}_8\times {(F_4)}_{10}$ is gaugeable, which makes the usual S-duality tricks useless for discerning that this is a product SCFT.}. Since it has rank-3, it must be a product of a rank-1 and a rank-2 theory. The possibilities for rank-1 $\mathcal{N}=2$ SCFTs are very limited \cite{Argyres:2015ffa,Argyres:2015gha,Argyres:2016xua,Argyres:2016xmc}. The only one consistent with the global symmetries and R-charges of the Coulomb branch parameters is the Minahan-Nemeschansky ${(E_7)}_8$ SCFT. The other factor in the product is, then, a new rank-2 SCFT, with global symmetry ${(F_4)}_{10}\times U(1)$, $n_4=n_5=1$ and $(n_h,n_v)=(32,16)$. So far, we are not aware of an alternative class-S construction of this theory.

\section*{Acknowledgements}
We would like to thank Philip Argyres, Chris Beem, Mario Martone, Leonardo Rastelli and Ken Intriligator for helpful discussions. This work was supported in part by the National Science Foundation under Grant No. PHY-1620610. J.D.~and F.Y.~would like to thank the Aspen Center for Physics, which is supported by National Science Foundation grant PHY-1066293,  for hospitality during the workshop ``Superconformal Field Theories in $d \geq 4$," where many of these ideas were clarified. They would also like to thank the organizers of the Pollica Summer Workshop 2017, partly supported by the ERC STG grant 306260, for their gracious hospitality during a later stage of this work.  Finally, J.D.~would like to thank the organizers of the KIAS Autumn Symposium on String Theory, for the opportunity to present these results prior to publication.

\bibliographystyle{utphys}
\bibliography{ref}

\end{document}